\documentclass[12pt]{article}

\usepackage{scicite}

\usepackage{times}
\usepackage{amsmath}
\usepackage{graphicx}
\usepackage{amssymb}
\usepackage{multibib}
\usepackage{color}
\usepackage{threeparttable}

\newcommand{\Fg}[1]{Figure~\ref{fig:#1}}

\newcommand{\Fgs}[2]{Figures\ \ref{fig:#1} and \ref{fig:#2}}

\newcommand{\eq}[1]{Eq.~(\ref{eq:#1})}
\newcommand{\Eq}[1]{Equation~(\ref{eq:#1})}

\newcommand{\eg}{e.g.}

\newcommand{\AU}{ \  \rm AU}
\newcommand{\Ms}{ \  \rm M_\odot }

\newcommand{\Me}{ \ \rm M_\oplus}
\newcommand{\taus}{ \tau_{\rm s}}
\newcommand{\tausF}{ \tau_{\rm s, F}}
\newcommand{\tausB}{ \tau_{\rm s, B}}

\newcommand{\Omegak}{\Omega_{\rm K}}

\newcommand{\alphat}{\alpha_{\rm t}}
\newcommand{\Mp}{M_{\rm p}}
\newcommand{\tauv}{\tau_{\nu}}

\DeclareMathSymbol{\varOmega}{\mathord}{letters}{"0A}
\DeclareMathSymbol{\varPhi}{\mathord}{letters}{"08}
\DeclareMathSymbol{\varSigma}{\mathord}{letters}{"06}
\DeclareMathSymbol{\varPsi}{\mathord}{letters}{"09}
\DeclareMathSymbol{\varGamma}{\mathord}{letters}{"00}

\newcounter{lastnote}

\date{}

\begin{document} 

\baselineskip24pt

\noindent {\bf Title} \\
Natural separation of two primordial planetary reservoirs in an expanding solar protoplanetary disk 

\noindent {\bf Authors} \\
Beibei Liu$^{1*,3}$  Anders Johansen$^{2,3}$, Michiel Lambrechts$^{2,3}$, Martin Bizzarro$^{2}$, \ and Troels Haugb{\o}lle$^{4}$

\noindent {\bf Affiliations} \\
$^{1}$ Zhejiang Institute of Modern Physics, Department of Physics \& Zhejiang University-Purple Mountain Observatory Joint Research Center for Astronomy, Zhejiang University, 38 Zheda Road, Hangzhou 310027, China \\
$^{2}$Center for Star and Planet Formation, GLOBE Institute, University of
Copenhagen, \O ster Voldgade 5-7, 1350 Copenhagen, Denmark \\
$^{3}$Lund Observatory, Department of Astronomy and Theoretical Physics, Lund University,  Box 43, 22100 Lund, Sweden\\
$^{4}$Niels Bohr Institute, University of Copenhagen, {\O}ster Voldgade 5-7, 1350 Copenhagen, Denmark

\noindent {\bf Abstract} \\
Meteorites display an isotopic composition dichotomy between non-carbonaceous (NC) and carbonaceous (CC) groups, indicating that planetesimal formation in the solar protoplanetary disk occurred in two distinct reservoirs. The prevailing view is that a rapidly formed Jupiter acted as a barrier between these reservoirs. We show a fundamental inconsistency in this model: if Jupiter is an efficient blocker of drifting pebbles, then the interior NC reservoir is depleted by radial drift within a few hundred thousand years. If Jupiter lets material pass it, then the two reservoirs will be mixed. Instead, we demonstrate that the arrival of the CC pebbles in the inner disk is delayed for several million years by the viscous expansion of the protoplanetary disk. Our results support that Jupiter formed in the outer disk (${>}10$ AU) and allowed a considerable amount of CC material to pass it and become accreted by the terrestrial planets.

\noindent {\bf \large Introduction}

The chronological ages of meteorites are inferred either by dating the formation of their main components (\eg , chondrules), or by dating  specific physicochemical processes (\eg, core-mantle segregation or aqueous alteration) to derive the parent-body accretion time through thermal modeling.  Such studies indicate that the accretion of meteorite parent-bodies in the Solar System started already within a few hundred thousand years from the formation of the CAIs\cite{Connelly2012} (generally taken to be the starting time of Solar System formation) and lasted for several Myr\cite{Amelin2002,Kleine2009,Kruijer2014}. The difference in composition between the NC group, originating in the inner Solar System, and the CC group, originating in the outer Solar System, needs to be present in the very early disk phase, within a few $10^{5}$ yr to explain the formation of primitive achondrites and iron meteorites of NC composition. These two reservoirs must remain radially separated at least for $2{-}3$ Myr in order for the primitive NC and CC parent-bodies to form around that time with distinct isotopic compositions\cite{Kruijer2017}.

Disk thermal processing has been proposed to explain the origin of these two disk reservoirs\cite{Trinquier2009,Ek2020}. In this scenario, Solar System dust is made up of refractory presolar grains and more volatile mantles grown in the interstellar medium (ISM)\cite{Zhukovska2008}. The presolar grains are s-process dominated and likely originate from asymptotic giant branch stars, whereas the ISM grains have a near-solar isotopic composition (CI-like) condensed from the gas that is already well-mixed from different isotopic sources (\eg, supernovae). The NC composition is then a result of the selective thermal destruction of either ISM-grown mantles or fragile presolar carriers\cite{Trinquier2009,Ek2020}. During the early infall phase,  the episodic accretion associated with an FU Orionis-type outburst temporally raises the disk temperature and moves the water ice line to the $20{-}30$ AU disk region\cite{Pignatale2018}. This would have naturally resulted in the separation of the NC component from the bulk CI component out to those distances, provided that the supernova-derived material resides chiefly in the fragile mantle of ice and organics. Alternatively, the initial two distinct disk reservoirs could also be inherited from isotopically heterogeneous accretion of molecular cloud gas onto the disk\cite{Nanne2019,Burkhardt2019}. 

Previous studies have suggested that an early and rapid growth of Jupiter is necessary to explain the subsequent $2{-}3$\,Myr separation of these two distinct  reservoirs\cite{Kruijer2017,Desch2018,Nanne2019}. Here, we find a fundamental inconsistency with the Jupiter barrier scenario, which requires an alternative model. We demonstrate that, after the formation of the protoplanetary disk, the combination of viscous spreading and radial drift of dust particles can consistently explain the nucleosynthetic isotopic compositions in both the asteroid and planetary record -- without invoking Jupiter to block the inwards flow of solid material.

\noindent {\bf \large Results}

 We provide a simple estimation of the solid particles' radial drift timescale.  Their radial drift velocity is given by (See Materials and Methods)
\begin{equation} 
   v_{\rm drift} = -  2 \taus \eta v_{\rm K}  + u_{\rm g}, 
   \label{eq:peb_drift}
\end{equation}
where $v_{\rm K}{\equiv} \Omega_{\rm K} r $ is the Keplerian velocity, $r$ is the orbital distance, $\Omega_{\rm K}$ is the angular velocity, $\eta {=} - \frac{1}{2}  \frac{\partial \ln P }{\partial \ln r} \left(\frac{H_{\rm g}}{ r} \right)^2$ is the headwind prefactor (a measure of how large the gas rotation velocity deviates from the Keplerian velocity), $P$ is the gas pressure, $H_{\rm g}$ is the disk gas scale height and $u_{\rm g}$ is the gas velocity at the disk midplane, which is proportional to the midplane turbulent viscous parameter $\alpha_{\rm t}$. The Stokes number of the particle is defined as $\taus{\equiv} \Omega_{\rm K} t_{\rm stop}$ where $t_{\rm stop}$ is the stopping time (a measure of how fast the particle adjusts its momentum to the surrounding gas). 

The particle's Stokes number in the fragmentation-limit growth regime is
\begin{equation} 
  \tau_{\rm s,F} =  \frac{v_{\rm F}^2}{\alpha_{\rm t} c_{\rm s}^2} \simeq 2\times  10^{-3}  \left(\frac{\alpha_{\rm t}}{2 \times 10^{-3}} \right)^{-1} \left(\frac{v_{\rm F}}{1.5 \rm \ m/s} \right)^2    \left(\frac{c_{\rm s}}{0.8 \rm \ km/s} \right)^{-2},
\end{equation}
where $v_{\rm F}$ is the fragmentation threshold velocity.
In the inner disk regions with a moderate midplane turbulence (\eg , at $5$ AU), $\alpha_{\rm t} {\lesssim} \tau_{\rm s,F}{\ll}1$, 
and the drift timescale can approximately expressed as  \begin{equation} 
  t_{\rm drift} \simeq \frac{1}{2 \eta \taus \Omega_{\rm K}} \simeq  0.06 \left( \frac{\eta}{0.007} \right)^{-1}  \left(\frac{\alpha_{\rm t}}{2 \times 10^{-3}} \right) \left(\frac{v_{\rm F}}{1.5 \rm \ m/s} \right)^{-2}    \left(\frac{c_{\rm s}}{0.8 \rm \ km/s} \right)^{2}  \left( \frac{r}{5 \ \rm AU} \right) \ \rm Myr.
\end{equation}  
The above timescale strongly depends on the fragmentation threshold $v_{\rm F}$. For a high fragmentation threshold ($v_{\rm F} {\gg} 1 \, {\rm m/s})$, the larger particle sizes lead to more rapid orbital decay of drifting particles, which results in a faster mixing between NC and CC reservoirs.  On the other hand, for disks with a high turbulent level $\alpha_{\rm t}{\gg}\tau_{\rm s,F}$, solid particles are well coupled to the gas and their radial drift is determined by the gas accretion flow.

We construct a $1$-dimensional model for the gas and solid components in a protoplanetary disk.  The disk is evolved using the conventional $\alpha$-viscosity approach. Most of the gas in the expanding disk is accreted onto the central star while the rest spreads outwards to conserve the angular momentum\cite{LyndenBell1974}. The solid particles are allowed to grow to their fragmentation-limited size \cite{Birnstiel2012}, yielding particle sizes that are smaller further from the star and later in time. The effects of the gas spreading and dust radial drift are included in the calculations.

We show in \Fg{migration}a that the particles initially at $r{\gg}r_{\rm exp}$ move outward together with the expanding gas, where $r_{\rm exp}$ is the separation radius between the gas inflow and outflow. Their motion then becomes inward when the particles’ radial drift overwhelms the outward spreading of the gas at later times. We find that dust particles starting at $1{-}10$ AU drift into the inner disk within $1$ Myr, whereas those initially located in the $20{-}30$ AU region sweep the asteroid belt $2{-}4$ Myr later, ending up with a millimeter characteristic size. This is also the case when we consider a disk with less vigorous turbulence in the midplane than the upper surface layer. In this so-called dead zone model the local midplane turbulence further depends on the column density of the gas in the protoplanetary disk (see Materials and Methods). We additionally account for the early disk building phase in \Fg{migration}c, where the angular momentum is dominated by the infall process with gas only being accreted inward. After a typical infall time of $0.3$ Myr, the disk viscosity dominates angular momentum transport and leads the outward spreading of the gas. Our result implies that the inner disk can preserve its NC nature until the CC dust particles drift through, as long as the initial boundary of NC/CC reservoirs ($r_{\rm CC}$) is larger than the original viscous expansion radius of gas ($r_{\rm exp}$).

In the prevailing view, the early and rapid formation of Jupiter has been invoked to explain the NC-CC separation \cite{Kruijer2017,Desch2018,Nanne2019}. When the core of Jupiter grows to approximately $10{-}20 \Me$ \cite{Lambrechts2014}, it starts carving a gap in the gas disk, preventing the outer CC solids from drifting inwards and mixing with the inner NC reservoir. Importantly, this model requires proto-Jupiter to form early within $0.5{-}1$ Myr in order to explain the formation of differentiated planetesimals in both reservoirs. We identify an inconsistency in this model in \Fg{dichotomy}a: if Jupiter appears early in the $5$ AU region and totally blocks the flow of solids, then the solids of NC composition interior of Jupiter would drift into the Sun within a few hundred thousand years (\Fg{dichotomy}a). This would then preclude the formation of the primitive NC parent-bodies after $2{-}3$ Myr.

Thus, a second scenario is considered in \Fg{dichotomy}b where Jupiter allows material to permeate past the gap and replenish the inner reservoir. However, in this circumstance the CC reservoir quickly overtakes the inner regions of the protoplanetary disk and there would be no compositional dichotomy preserved in the primitive NC/CC parent bodies formed after 2--3 Myr. Reducing or stopping the radial drift of solids interior of Jupiter could solve this inconsistency\cite{Kruijer2017,Desch2018}, but this suggestion comes at the expense of ignoring or reducing the radial drift of solids -- a fundamental physical process operating in protoplanetary disks -- in order to separate the reservoirs.

 In this context, it has then been suggested that the NC/CC reservoirs were isolated due to trapping of pebbles in speculative long-lived pressure bumps in the inner disk\cite{Brasser2020} or through continuous and efficient planetesimal formation around the water ice line\cite{Lichtenberg2021,Charnoz2021}. The latter model requires that the inner disk experiences either intensive irradiation or  strong viscous heating to keep the water ice line away from the terrestrial planet zone and  additionally extremely high planetesimal formation efficiencies that convert hundreds of Earth masses of CC material to planetesimals. This raises the concern that a large fraction of those planetesimals would subsequently become implanted in the asteroid belt due to the migration of Jupiter\cite{Pirani2019}.

Here we instead propose that the viscous expansion of the gas naturally separates the two disk reservoirs (\Fg{dichotomy}c). As such, there is no need to invoke very early Jupiter formation in order to explain the NC-CC separation.
In this case, the core of proto-Jupiter forms in a wide orbit and then undergoes inward migration.
Adopting a state-of-the-art pebble accretion and planet migration model\cite{Liu2019},  we provide an illustration of Jupiter's growth and migration track in \Fg{dichotomy}c. In our model, Jupiter finishes the growth of its $10 \Me$ core at a time of $t{\simeq}3$ Myr in the $10$ AU disk region and reaches its full mass (by gas accretion) and the current position at $t{\simeq}4.2$ Myr.

These two contrasting models for the NC/CC dichotomy in the context of the known constraints on parent-body formation timings are illustrated in \Fg{timetable} (see also Materials and Methods). The expansion of the protoplanetary disk is a key physical mechanism needed to understand the asteroid and planetary record of the Solar System.
As shown in \Fg{dichotomy}c, with viscous disk spreading, the outer CC material arrives in the inner disk region only after $2{-}3$ Myr. 
This expansion allows the parent bodies of NC differentiated meteorites, achondrites, enstatite, and ordinary chondrites to form sequentially during the first $2{-}3$ Myr in the inner disk region (\Fg{timetable}c). The later formed ordinary chondrites reveal a modest compositional deviation from earliest formed Ureilites towards CC chondrites, implying a continuous admixing between the original inner NC disk reservoir and outer CI-like material (\eg , ureilites and CI chondrites as end members for the mixing)\cite{Schiller2018}. This is a natural consequence of the planetesimals forming at the time where the NC and CC reservoirs partly mix, as illustrated by the shaded region in \Fg{timetable}c. The small (${<}100$ km) NC asteroids are protected against further accretion of substantial amounts of drifting CC pebbles. The pebble accretion efficiency depends strongly on the inclinations and eccentricities of the planetesimals. The internal gravitational stirring of the primitive NC asteroids will quickly raise their eccentricities to a degree where subsequent accretion of CC material is insignificant (see Materials and Methods). A similar early excitation of the orbit of Mars has been proposed to explain how Mars terminated its accretion earlier than Earth and thus incorporated a lower fraction of CC material\cite{Schiller2018}.

The differentiated meteorite parent-bodies of CC composition form at relatively large orbital distances ($r{>}r_{\rm CC}$) within $1$ Myr, whereas the parent-bodies of CC chondrites form at $t{\simeq}3{-}4$ Myr exterior to the orbit of Jupiter and are later scattered into the asteroid belt by the combined gravity of Saturn and Jupiter\cite{Eriksson2020}. The terrestrial planets -- proto-Earth and Mars -- accrete considerable CC material and complete their main accretion within $4{-}5$ Myr\cite{Warren2011, Schiller2018,Johansen2021}. Our model allows Jupiter'core to finalize its growth at a relatively late time (\eg , $t{\sim}2{-}3$ Myr) and at large orbital distances, such as $r{>}10$ AU, in agreement with planet formation models that include inwards migration due to gravitational torques exerted by the protoplanetary gas disk on the planet\cite{Coleman2014, Bitsch2015}. Large-scale inward migration of proto-Jupiter plays a decisive role in sweeping through the formation zone of CC differentiated planetesimals. The orbital crossings and close-encounters between Jupiter and those planetesimals result in the CC iron meteorite parent-bodies to be frequently scattered into the inner asteroid belt region\cite{Raymond2017}.

\noindent {\bf \large Discussion}

We first discuss the role of Jupiter on dust filtration. Hydrodynamic studies find that while large mm-to-cm size particles are stopped at the pressure maximum exterior to the planet, small grains that are well coupled to the disk gas can pass the gap\cite{Zhu2012, Weber2018,Haugbolle2019, Drazkowska2019}. How the above size-dependent filtering affects the isotope anomalies is not consistently accounted for in the early Jupiter scenario\cite{Kruijer2017}. We show in \Fg{filtering} the efficiency of Jupiter as a dust filter, taking into account a size distribution of solids. Two different viscosity prescriptions are considered: a constant $\alpha$ everywhere in the disk (blue) and a higher value exterior to the gap (red). The latter assumes an elevated viscosity due to the generation of vortices by the Rossby wave instability associated with gap-opening in the inner regions of the disk where the background turbulence is inherently low\cite{Hallam2017}. The critical Stokes number of particles to cross the gap decreases as Jupiter's mass increases, whereas the maximum Stokes number of particles in the fragmentation regime remains constant at the planet location. When the mass of Jupiter is low, all particles are below the critical Stokes number and can drift through the gap. Jupiter begins to block particles only when the critical Stokes number is lower than the fragmentation-limited particle size.  Although the maximum size particles dominate the total mass of the whole dust population, a non-negligible fraction of them would unavoidably fragment into smaller particles. We assume that the dust size distribution follows a power law, $n(a) \propto a^{p}$, where $p{=}{-}3.5$, $a_{\rm min}{=} 0.1 \rm \ \mu m$ and $a_{\rm max}$ is the maximum size in the fragmentation-limit. We find that a dust fraction of $(a_{\rm crit}/a_{\rm max})^{4-p}$  still diffuse through the gap region when $a_{\rm max}{>}a_{\rm crit}$. 
As shown in \Fg{filtering}e when Jupiter grows rapidly, $24\%$ and $65\%$ of CC dust would drift through Jupiter's gap after $5$ Myr at $\alpha_{\rm gap}{=}0.001$ and $0.003$. This dust filtering works much less effectively for a $3$ Myr gas accretion timescale in \Fg{filtering}f. 

We note that larger cm-sized CAIs, condensed in the inner regions of the protoplanetary disk and likely transported to the outer disk by winds and outflows, will be efficiently trapped by the growing Jupiter and their igneous nature will prevent the formation of CAI fragments that could pass the gap and hence this can explain how the ordinary chondrites avoided accreting CAIs\cite{Desch2018, Haugbolle2019}.

Our scenario supports that the disk angular momentum is transported outwards by the turbulent viscosity, for instance, via the magnetorotational instability (MRI)\cite{Balbus1991}.  Although MRI turbulence is likely to be suppressed due to non-ideal MHD effects in the inner ten AU disk region where disk winds could drive a laminar accretion flow, it can operate at large disk radii where the CC dust particles reside initially.  In the outer disk region where ambipolar diffusion dominates, the turbulent level depends on the strength of the net magnetic field, dust abundance and XUV/cosmic ray penetration depth\cite{Simon2015}. The disk turbulence can be inferred from the spectral energy distribution or molecular line velocity measurements. Such observations show that some disks display a relatively high $\alpha{\gtrsim}10^{-3}{-}0.1$\cite{Ovelar2016,Flaherty2020}, compatible with measurements from  non-ideal MHD disk simulations\cite{Simon2015}.  Moreover, a recent study found that even in the magnetized wind-driven disk where the MRI is fully quenched, the magnetic torque can also drive the expansion of the gas in the outer disk region,  analogous to the viscous spreading effect\cite{Yang2021}. This implies that our expanding disk scenario can be applicable for a broad range of circumstances. Despite uncertainties in measuring disk size,  the observed sizes of Class II disks are plausibly larger than those of Class I disks, in alignment with the above theoretical  predictions\cite{Najita2018}.

\noindent {\bf \large Summary}

In summary, we propose that a generic outward-then-inward radial motion of pebbles, driven first by the viscous expansion of the disk gas and thereafter by the radial drift of pebbles, is the key ingredient that explains the isotopic compositions of the entire known meteoritic and planetary record. In this view, the observed NC/CC dichotomy is set by a secular change to the composition of the disk and not by an early spatial separation by an in-situ formed Jupiter\cite{Kruijer2017, Desch2018}. We show that the gap opening by Jupiter will not stop all inward-drifting solid particles and that Jupiter lets considerable amounts of CC dust and pebble fragments enter the inner disk to contribute to the mass budgets of the terrestrial planets as well as to the parent bodies of the ordinary chondrites. 

Our work implicitly supports a distant origin of Jupiter,  which is consistent with orbital migration theory\cite{Coleman2014, Bitsch2015,Pirani2019}. 
Indeed, many heavy elements including ultra-volatile noble gases are found to be enriched in Jupiter’s atmosphere compared to the solar photospheric values. This feature cannot be explained by in-situ accretion and therefore strongly favors that Jupiter's core formed in a distant, cold orbit (\Fg{timetable}c)\cite{Oberg2019}. In this view, heavy elements can be accreted as solids that would later become diluted into the envelope, resulting in the observed enrichment pattern.

\noindent {\bf \large Main Figures} \\

\begin{figure}[!ht]
\includegraphics[width=16cm]{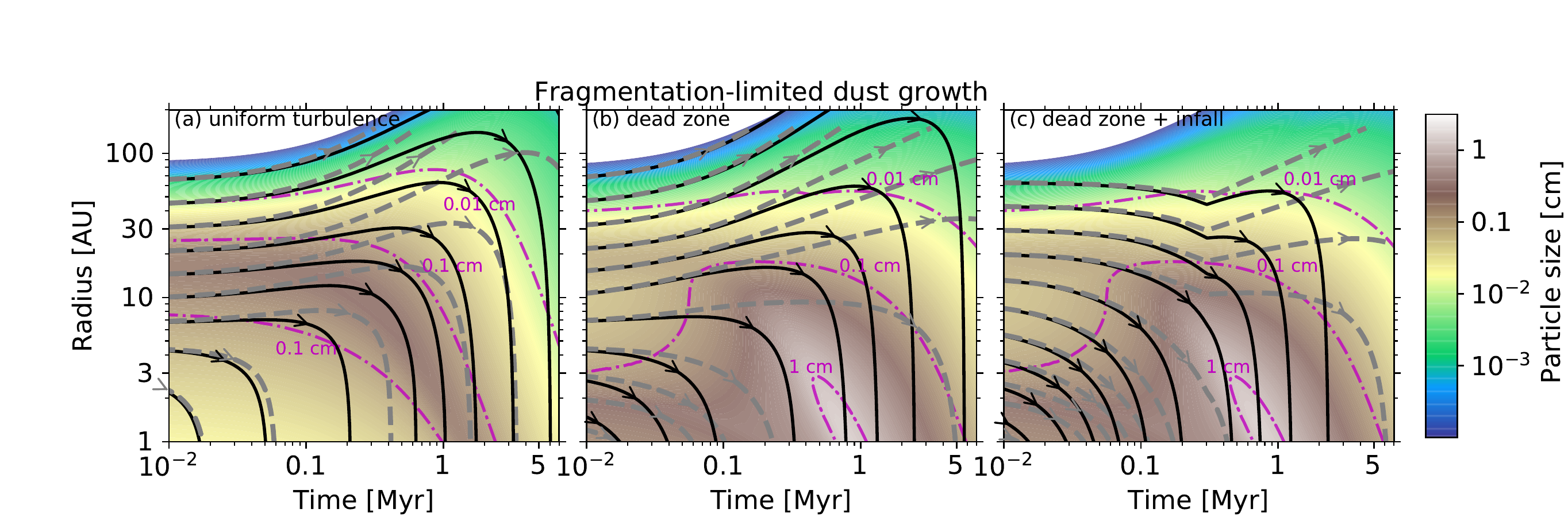}
\caption{\textbf{Radial drift of dust particles in a viscously spreading disk with (a) uniform turbulent model, (b) dead zone model, and (c) dead zone + infall model.}
The black curves trace the radial drift of dust particles whose size is limited in the fragmentation regime,  while the grey dashed lines indicate the gas trajectories. The background color refers to the size of the particle at the corresponding time and disk location, with the magenta dash-dot lines marking pebbles of $0.1$, $1$ and $10$ mm in size. The disk parameters are: $v_{\rm F}{=}1 \rm \ m/s$, $r_{\rm d}{=}10$ AU, $C_{\rm d}{=}0.14 \ M_{\odot}$,   $T_{\rm irr0}{=} 130$ K and  $\alpha{=}5{\times }10^{-3}$ in the uniform turbulent model, $\alpha_{\rm A}{=}10^{-2}$ and $\alpha_{\rm D}{=}5{\times} 10^{-4}$ in the dead zone model (fiducial run). In the dead zone + infall model, the early infall period where gas only accretes inward lasts for $0.3$ Myr, and after that, the disk gas transports angular momentum through viscous spreading. The outwards flow of gas in the expanding disk brings along with it the solids from the outer regions of the disk. As the solids detach from the gas flow, they drift into the inner disk regions with the arrival time determined by their initial distance
from the star. } 
\label{fig:migration}
\end{figure}

\begin{figure}[!ht]
\includegraphics[width=16cm]{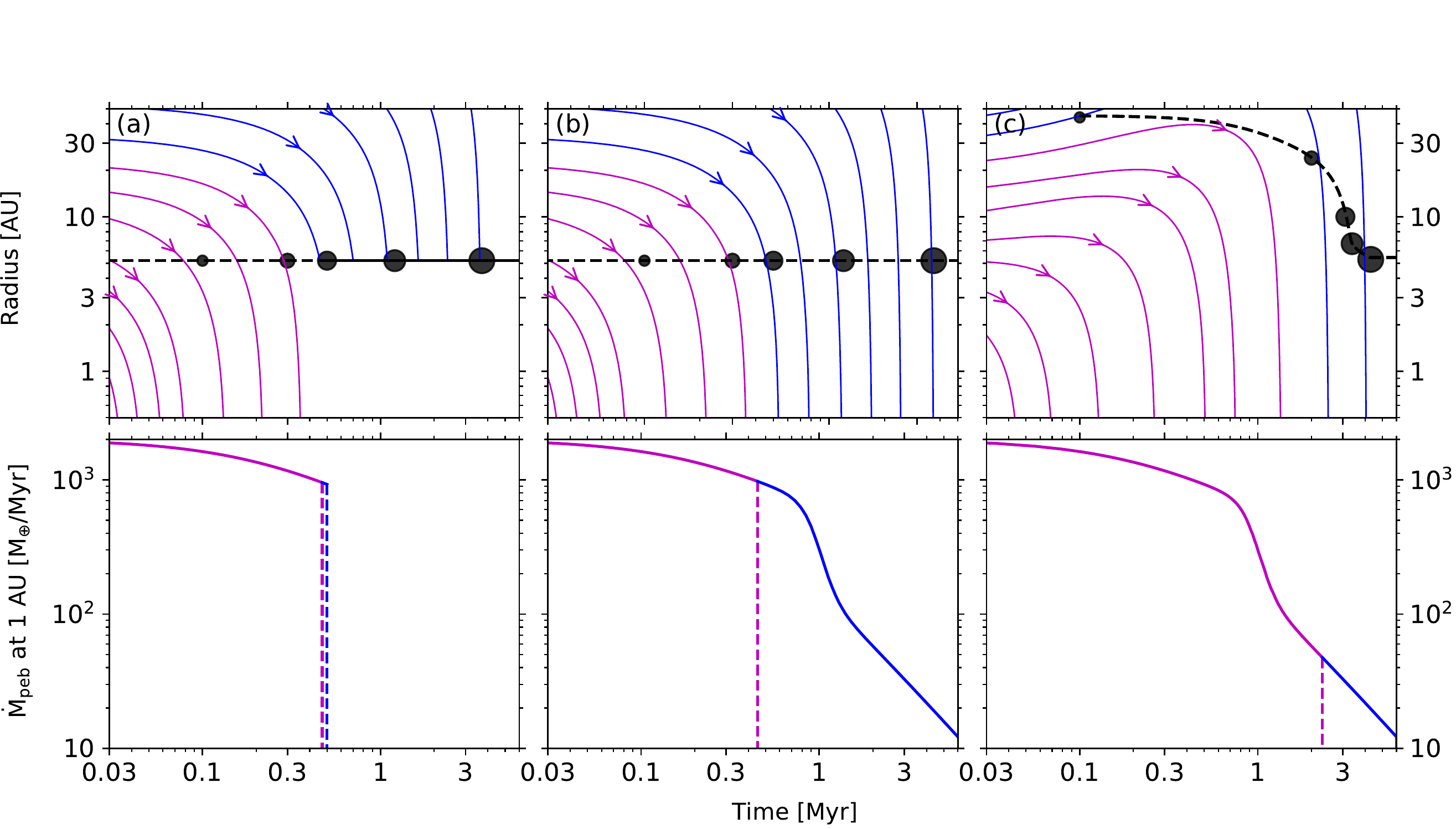}
\caption{\textbf{Radial drift of pebbles (top) and the pebble flux entering the inner terrestrial planet formation region (bottom).}
The trajectories of the NC and CC dust particles are shown in red and blue, while the growth of Jupiter is traced by the black curve with $1 \Me$, $3\Me$, $10\Me$, $100 \Me$ and full mass in circles. The gas moves only inwards in panels (a) and (b) whereas the gas outward viscous spreading is included in panel (c). In panel (a) we assume that Jupiter formed early and that it entirely blocked the inward drifting solid particles after $0.5$ Myr, while in panel (b) Jupiter is assumed to let all dust particles pass through. In the first case, the inner (NC) reservoir is quickly depleted, while in the second case the CC reservoir quickly takes over the inner disk. In panel (c) we show that viscous expansion can keep the two disk reservoirs separated within $2{-3}$ Myr. Jupiter is here assumed to form at a wide orbit after $1$ Myr.
In top panels the black dashed lines represent Jupiter is not an efficient dust filter while the black solid line assumes that solid particles are all blocked by Jupiter. The vertical dashed lines in the bottom panels mark when the NC and CC reservoirs get truncated. For example, in the left bottom panel the NC reservoir is depleted at $t{=}0.45$ Myr while Jupiter blocks the CC reservoir at $t{=}0.5$ Myr. 
The disk parameters are: $v_{\rm F}{=}1.5 \rm \ m/s$, $r_{\rm d}{=}8$ AU, $C_{\rm d}{=}0.14 \ M_{\odot}$,   $T_{\rm irr0}{=} 130$ K, $\kappa_0{=}10^{-2} $ and  $\alpha_{\rm A}{=}8\times 10^{-3}$ and $\alpha_{\rm D}{=}10^{-3}$.} 
\label{fig:dichotomy}
\end{figure}

\begin{figure}[!ht]
\includegraphics[width=10cm]
{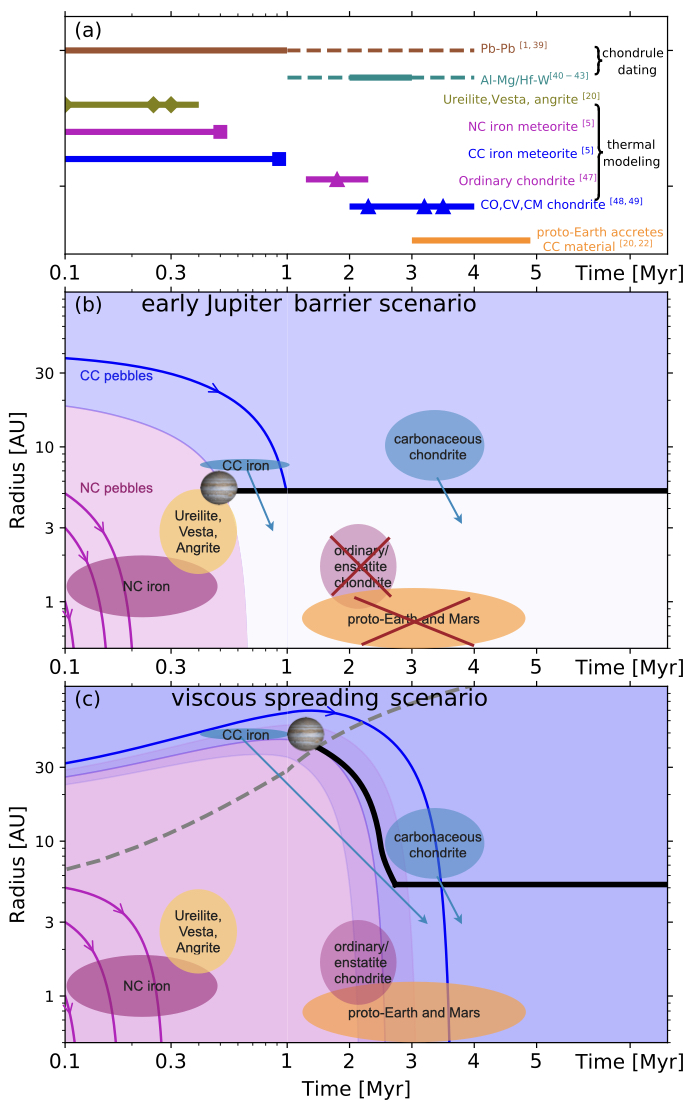}
\caption{\textbf{Meteorite dating records and a schematic illustration of the two Solar System formation scenarios.} 
Panel (a): symbols give the peak accretion ages of meteorites with uncertainties shown in horizontal lines. Panel (b,c): The magenta and blue lines show example trajectories of NC and CC dust particles, while the dashed line in the viscous spreading scenario is the separation radius for gas inflow/outflow. For the early Jupiter scenario, the core grows in-situ within $1$ Myr and blocks all particles drifting from the outer part of the disk. The inner disk drains out solids quickly, which conflicts with the later formation of ordinary/enstatite chondrites and the incorporation of CC material into the terrestrial planets. For the viscous spreading scenario, Jupiter's core forms beyond (and slightly later than) the CC differentiated parent bodies. The parent bodies of primitive NC meteorites (ordinary chondrites and enstatite chondrites) form at $2{-}3$ Myr while the inner disk region is changing composition.  The primitive CC parent bodies form at $t{\sim}3{-}4$ Myr and are scattered into the asteroid belt together with the CC differentiated parent bodies by Jupiter's migration.}
\label{fig:timetable}
\end{figure}

\begin{figure}[!ht]
\includegraphics[width=10cm]
{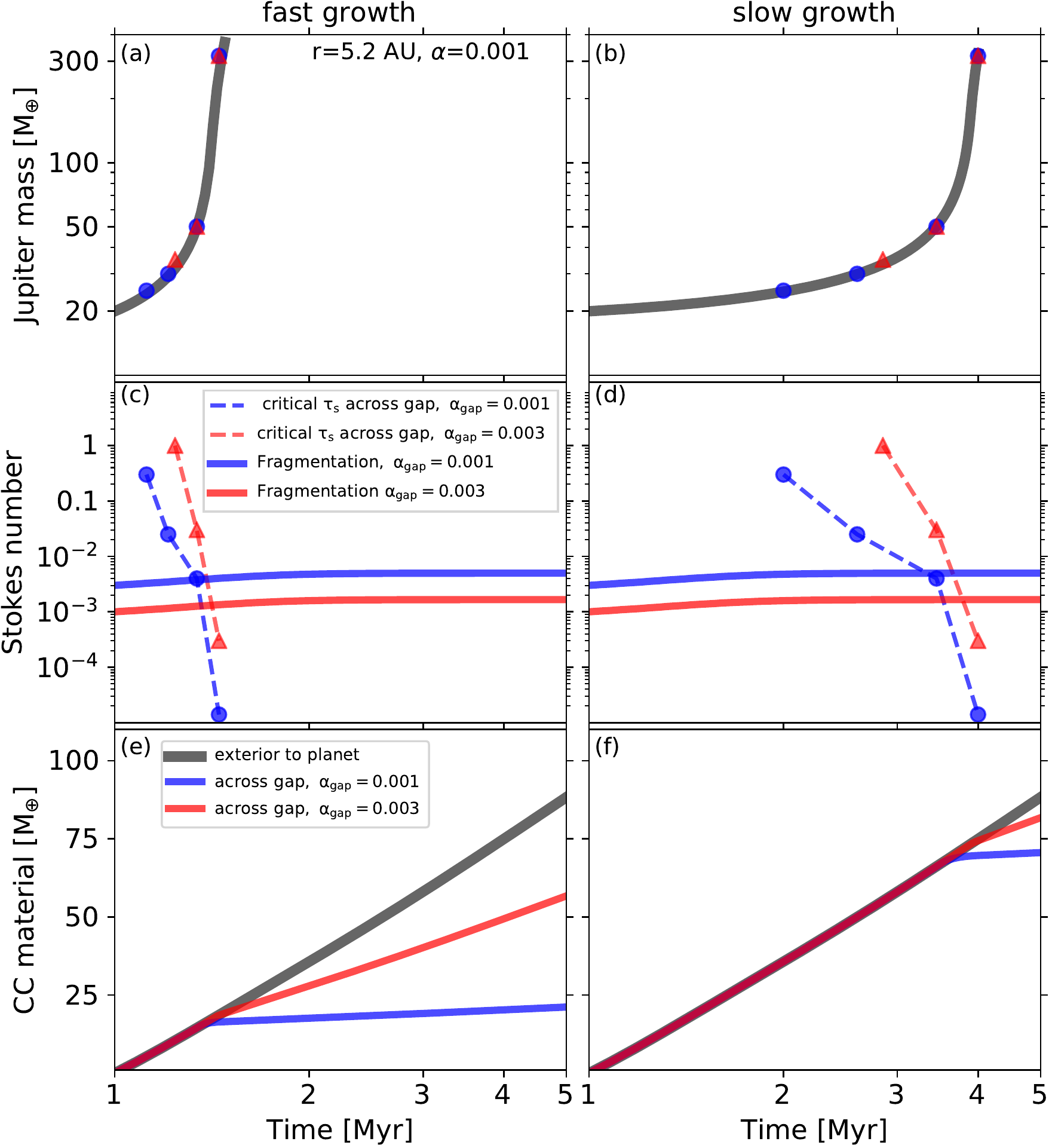}
\caption{\textbf{Evolution of Jupiter's mass, particle Stokes numbers and CC dust mass interior to Jupiter after its core formation.}
Panel (a+b): the mass of Jupiter starting from $20 \Me$ at $t{=}1$ Myr, with a characteristic growth
timescale of $0.5$ Myr (left) and $3$ Myr (right).  Panel (c+d): critical (maximum) Stokes number $\tau_{\rm s, crit}$ that particles can diffuse through the gap vs disk particles' Stokes number in the fragmentation regime for $\alpha_{\rm gap}{=}0.001$ (blue) and $0.003$ (red). The blue dots and red triangles refer to $\tau_{\rm s, crit}(M_{\rm p})$ at two different $\alpha_{\rm gap}$, where the data from the low-masses and Jupiter's full mass are adopted from Ref\cite{Bitsch2018} and setup I of Ref \cite{Haugbolle2019}, respectively. From Ref\cite{Haugbolle2019} we calculate the critical sizes when the dust-to-gas ratio interior of the planet drops below 50\% of the exterior value and convert those into Stokes numbers.  Panel (e+f): The mass of CC dust particles that drift across the Jupiter's gap for for $\alpha_{\rm gap}{=}0.001$ (blue) and $0.003$ (red). The grey line shows the CC dust drift exterior to the orbit of Jupiter for comparison.  The other disk parameters are $\alpha {=}10^{-3}$, $T_{\rm irr0}{=}130$ K, $C_{\rm d}{=}0.14 \ M_{\odot}$, $v_{\rm F} {=} 1 \ \rm m/s$ and $r_{\rm d}{=}10$ AU. The growing Jupiter lets a significant fraction of the dust mass past its orbit, unless it grows very fast and has a low turbulent diffusion outside the gap.} 
\label{fig:filtering}
\end{figure}

\clearpage

\noindent {\bf \large Materials and Methods}

\noindent {\bf Chronology of meteorites}

We summarize the literature chronological meteorite studies in \Fg{timetable}a. The ages of chondritic parent-bodies can be deduced from the time when chondrules were last melted and crystallized. This age is determined from either the long-lived, U-corrected Pb-Pb dating or short-lived Al-Mg and Hf-W datings. In the Pb-Pb dating approach, recent studies proposed that chondrules form simultaneously with CAIs and last for $3{-}4$ Myr, whereas the age peaks around $0$ to $1$ Myr\cite{Connelly2012, Bollard2017}. In the Al-Mg dating approach, chondrules typically record a lower initial Al isotopic ratio compared to that of CAIs, the latter of which is treated as the canonical value for the onset of the Solar System ($^{26}$Al/$^{27}$Al ${=}5.5\times 10^{-5}$). This is interpreted as a $1{-}4$ Myr protracted formation of chondrules after CAIs, with a major age peak at $t{=}2{-}3$ Myr\cite{Villeneuve2009, Kita2012,Budde2018, Pape2019}. Note that Pb-Pb dating of single chondrules shows different ages from that of the pooled chondrule separates, the latter of which seems more aligned with the results obtained from the Al-Mg dating \cite{Kruijer2020}. On the other hand, the short-lived Al-Mg dating relies on the assumption of an initial homogeneous distribution of $^{26}$ Al throughout the inner Solar System, the validity of which is subject to debate in the community\cite{Schiller2015, Bollard2019}. 

On the other hand, the age of meteorites can instead be inferred from the isotope dating of specific physical/chemical events occurred in their history. For iron meteorites (differentiated planetesimals), Hf-W dating probes the ages when their parent-bodies' cores became segregated from mantles. Then the formation time of these bodies can be traced back via thermal modeling of their internal evolution, based on the assumption that $^{26}$Al is the predominant heat source. Similar approaches have been used for achondrites, ordinary and carbonaceous chondrites, with different dating events such as magma crystallization, formation of minerals like carbonates, and secondary fatalities. In such a way, Vesta and the parent-bodies of ureilites, angrites are deduced to form within $0.5$ Myr after the CAI formation\cite{Schiller2018}. 
The parent-bodies of iron meteorites are found to form earlier than $0.5$ Myr for the NC population and earlier than $1$ Myr for the CC population\cite{Kruijer2017}. The parent-body accretion of ordinary chondrites takes place at $t{\sim}1.6{-}1.8$ Myr\cite{Doyle2015}, while those of carbonaceous chondrites occur at ${\sim}2.5$ to $4$ Myr after CAIs, spanning from CO, CV and CM types\cite{Fujiya2012, Jogo2017}. 

Furthermore, the accretion history of terrestrial planets can be constrained from their isotopic compositions. Many isotopes such as  $^{48}$Ca, $^{50}$Ti and $^{54}$Cr of Earth, Mars and Moon are found to be different from either NC or CC asteroid populations\cite{Warren2011, Schiller2018}. The terrestrial planets originally grew in the NC environment. One expects that these NC/CC disk reservoirs are temporarily separated early on and mixed together later through drifting of CC dust particles. As such, in order to account for their isotopic compositions, Earth and Mars need to accrete a non-negligible fraction of CC material (an upper limit of $30{-}40\%$ depending on different estimated isotopes) during the protoplanetary disk phase. Inferred from Hf-W chronometry, Mars attains its half mass within $3$ Myr\cite{Dauphas2011}, while the formation of Earth takes much longer, approximately at $30{-}100$ Myr\cite{Kleine2009}. The above Earth age derivation, however, could be substantially affected by the moon-forming impact. It is plausible that proto-Earth had completed the main accretion within the gas disk lifetime and the last Moon-forming giant impact happened a few tens of Myr later\cite{Yu2011}. The early and rapid formation of proto-Earth is more consistent with iron isotope evidence\cite{Schiller2020}.  Ref\cite{Johansen2021} proposed a pebble accretion terrestrial planet formation model that explains their architecture, bulk compositions as well as isotopic variability. They suggested that Mars and proto-Earth accrete CC materials $3{-}4$ Myr after the CAI formation.

\noindent {\bf Spreading disk model}

We want to track the $1$D evolution of the gas and solid components in the protoplanetary disk.  A conventional disk model is constructed, including both the gas spreading and dust radial drift.  We focus on the stage when the star completed its main accretion and disk has already formed. Thus, the disk-cloud interaction and infall of envelope material onto the disk are neglected for our fiducial setup. The influence of the early infall is further addressed in Section \textbf{Early disk formation}. Although the disk formation and evolution are found to be much more complicated in zoom-in $3$D hydrodynamic simulations\cite{Kuffmeier2017}, the advantage of our simplified $1$D treatment is that it captures the essential physical processes, at the same time with the feasibility of testing the influence of model parameters at a considerable detail.  Such explorations can inspire and guide future sophisticated numerical modelings.
 
We recapitulate the classical theory of viscous spreading disk\cite{LyndenBell1974,Hartmann1998}.   Based on the assumption that viscosity increases linearly with radial distance $\nu{ = }\nu_{\rm d} (r/r_{\rm d})$,  the characteristic timescale of viscous accretion/spreading is expressed as 
\begin{equation} 
  t_{\nu} = \frac{r_{\rm d}^2}{3 \nu_{\rm d}},
\end{equation} 
where $r_{\rm d}$ is the characteristic disk radius at $t{=}0$ yr, and $\nu_{\rm d}$ is the corresponding viscosity at $r_{\rm d}$. Here we define the dimensionless time as $\tauv {=} 1 + t/t_{\nu}$. The gas surface density and disk accretion rate from the self-similarity solution are written as\cite{LyndenBell1974}
\begin{equation} 
  \Sigma_{\rm g}= \frac{M_{\rm d0}}{2 \pi r_{\rm d}^2} \left( \frac{r}{r_{\rm d}} \right)^{-1} \tauv^{-1.5} \exp\left[ - \left( \frac{r}{r_{\rm d}}\right) \tauv^{-1}\right],
  \label{eq:disk1}
\end{equation}   
and  
\begin{equation} 
  \dot M_{\rm g}=- 3 \pi \Sigma_{\rm g} \nu\left[   1- 2 \left(\frac{r}{r_{\rm d}}\right) \tauv^{-1}  \right] = - \frac{M_{\rm d0}}{2 t_{\nu}}  \tauv^{-1.5} \exp\left[ - \left( \frac{r}{r_{\rm d}}\right) \tauv^{-1}\right] \left[   1- 2 \left(\frac{r}{r_{\rm d}}\right) \tauv^{-1}  \right],
  \label{eq:mdot}
\end{equation} 
where $M_{\rm d0}$ is the initial gas disk mass and $r$ is the radial distance from the central star. 
The gas velocity is given by 
\begin{equation} 
  u_{\rm g} =\frac{\dot M_{\rm g}}{ 2 \pi  r \Sigma_{\rm g}} =  -\frac{3 \nu}{2r}   \left[ 1- 2\left(\frac{r}{r_{\rm d}}\right) \tauv^{-1}\right].
  \label{eq:vgas}
\end{equation}

Hereafter we consider a realistic two-component disk structure based on different heating sources.  The inner disk region is heated by viscous dissipation while the outer disk region is dominated by stellar irradiation\cite{Liu2019}. The standard $\alpha$ prescription is adopted for viscosity $\nu{ =} \alpha R_{\rm g} T/ (\mu  \Omegak) $\cite{Shakura1973} where $R_{\rm g}$ is the gas constant,  $T$ is the disk temperature, $\mu$ is the gas mean molecular weight and $\Omegak{=}\sqrt{GM_{\star}/r^3}$ is the angular velocity and   $M_{\star}$ is the stellar mass. The self-similarity solution is applicable for the outer stellar irradiated disk region. The gas temperature and surface density are given by 
\begin{equation}
  T_{\rm irr} = \left( \frac{\phi L_{\star}}{4 \pi \sigma_{\rm SB}  \AU^2} \right)^{1/4}    \left( \frac{r}{1 \AU} \right)^{-1/2} = T_{\rm irr0} \left( \frac{r}{1 \AU} \right)^{-1/2}.
  \label{eq:T_irr}
\end{equation} 
\begin{equation} 
  \Sigma_{\rm g, irr} = \frac{C_{\rm d}}{2 \pi r_{\rm d} \AU}  \tauv^{-1.5} \exp\left[ - \left( \frac{r}{r_{\rm d}}\right) \tauv^{-1}\right]\left( \frac{r}{\AU} \right)^{-1},
  \label{eq:Sig_irr}
\end{equation}
where $\phi$ is the grazing angle where the stellar light hits the disk surface, $L_{\star}$ is the stellar luminosity, $C_{\rm d}$ is the equivalent disk mass when the disk is purely stellar irradiated from \eq{disk1},  $\sigma_{\rm SB}$ is the Stefan-Boltzmann constant and AU is the mean distance between Earth and Sun. It is worth noting that  $\phi$ is much lower than the unit and should principally depend on $r$ in a sophisticated disk model (leading to $T_{\rm irr}{\propto}r^{-3/7}$)\cite{Chiang1997}. Here we simply adopt it to be a constant.  For the fiducial disk model when $\phi{=}0.01$ and $L_{\star}{=}1 \  L_{\odot}$,  the irradiation temperature $T_{\rm irr0}$ at $1$ AU  is $130$ K.   Due to the variation of $\phi$ (related to disk opacity) and more luminous young star with a higher $L_{\star}$ during its pre-main sequence,  $T_{\rm irr0}$ could be varied from $130$ K to $300$ K.

For the inner viscously heated disk region, the viscosity depends on both the disk surface density and radial distance,  and hence,  the self-similarity solution is not exactly applicable.  We assume that in the inner disk region the gas flow is maintained by \Eq{mdot} such that $(\Sigma_{\rm g} \nu)_{\rm vis} {=} (\Sigma_{\rm g} \nu)_{\rm irr}$. In the inner viscously heated region,  the energy balance between heating and cooling provides that  
\begin{equation} 
  \frac{9}{4} \Sigma_{\rm g} \nu  \Omega_{\rm K}^2 = 2 \sigma T_{\rm e}^4,
  \label{eq:energy}
\end{equation}
where $T_{\rm e}$ is the effective gas temperature at the disk surface.  The heat that is generated from viscous dissipation in the midplane radiates vertically towards to the surface layer. For an optically thick radiative disk, the midplane temperature gives $T^4{\simeq} 3\tau /8 T_{\rm e}^4$ and the optical depth is $\tau {=} \kappa \Sigma_{\rm g}$.  We follow a simple  disk opacity law $\kappa{=} \kappa_0 T$\cite{Liu2019}, and then the energy equation reads 
\begin{equation} 
  \frac{9}{4} \Sigma_{\rm g} \nu  \Omega_{\rm K}^2 = \frac{16 \sigma T^3}{3 \kappa_0 \Sigma_{\rm g}}.
  \label{eq:energy2}
\end{equation}
We obtain 
\begin{equation} 
  T_{\rm vis} =T_{\rm vis0}  \tauv^{-3/4} \exp\left[ - \left( \frac{r}{2 r_{\rm d}}\right) \tauv^{-1}\right] \left( \frac{r}{1 \AU} \right)^{-9/8},
  \label{eq:T_vis}
\end{equation} 
and 
\begin{equation} 
  \Sigma_{\rm vis} =  \left(\frac{C_{\rm d}}{2 \pi  r_{\rm d} \AU}  \frac{T_{\rm irr0}}{T_{\rm vis0}} \right)   \tauv^{-3/4} \exp\left[ - \left( \frac{r}{2r_{\rm d}}\right) \tauv^{-1}\right] \left( \frac{r}{1 \AU} \right)^{-3/8},
 \label{eq:Sig_vis}
\end{equation}
where
\begin{equation} 
 T_{\rm vis0} =\left( \frac{27 \alpha \kappa_0 R_{\rm g} \Omega_{\rm K0}}{64 \mu \sigma_{\rm SB}} \right)^{1/4} \left(  \frac{C_{\rm d}}{2 \pi r_{\rm d} \AU} \right)^{1/2} T_{\rm irr0}^{1/2},
 \label{eq:T_vis0}
\end{equation}
and $\Omega_{\rm K0}$ is the Keplerian angular velocity at $1$ AU. 

The transition radius $r_{\rm tran}$ between these two disk regions can be solved by equating $T_{\rm vis}$ and $T_{\rm irr}$,  
\begin{equation} 
  r_{\rm tran}^{-5/8} \exp\left[ - \left( \frac{r_{\rm tran}}{2r_{\rm d}}\right) \tauv^{-1}\right] = \left(\frac{T_{\rm irr0}}{T_{\rm vis0}}\right)  \tauv^{3/4}
  \label{eq:rtran}
\end{equation}

The initial disk mass is
\begin{equation} 
  M_{\rm d0}= 2\pi \int_{\rm in}^{\rm out}  r  \Sigma_{\rm g}(t=0) d r,
\label{eq:mdisk}
\end{equation}
and the outer edge of the disk (or disk size, $r_{\rm out}$) defines that includes $95\%$ of its mass. The above two quantities can be solved numerically. For the fiducial setup in Table 1, the initial disk mass $M_{\rm d0}$ is $0.065 \ M_{\odot}$ for the uniform turbulence model and  $0.06 \ M_{\odot}$ for the dead zone model. \Fg{diskmodel} illustrates the evolution of gas temperature and surface density for our fiducial setup.

We note that the disk angular momentum is transported  from the inner region of the disk to the outer region through the viscous stress. As such,  most of the disk gas moves inward while the remaining gas expanse to larger radii to conserve angular momentum. Based on \eq{vgas}, the radius that separates mass inflow and outflow is expressed  as
\begin{equation} 
  r_{\rm exp} =  \frac{r_{\rm d}\tauv}{2} = \frac{r_{\rm d}}{2}   \left( 1 +\frac{ t}{t_{\nu}}  \right).
  \label{eq:rs}
\end{equation}
At the disk radius larger than $r_{\rm exp}$, the mass flux moves outward and $u_{\rm g}{>}0$; at the disk radius smaller than  $r_{\rm exp}$, the mass flux moves inward and $u_{\rm g}{<}0$.  Inserting \eq{rs} into \eq{vgas},  we rewrite the gas velocity as 
\begin{equation} 
  u_{\rm g} = -\frac{3 \nu}{2r}   \left( 1- \frac{r}{r_{\rm exp}}\right).
  \label{eq:vgas2}
\end{equation} 

The radial drift velocity of solid particle is given by\cite{Nakagawa1986} 
\begin{equation} 
   v_{\rm drift} = - \frac{2 \taus}{1 + \taus^2} \eta v_{\rm K}  + \frac{1}{1 + \taus^2}u_{\rm g}, 
   \label{eq:vpeb}
\end{equation}
where $v_{\rm K}{\equiv} \Omega_{\rm K} r $ is the Keplerian velocity.  The Stokes number of the particle $\taus{\equiv} \Omega_{\rm K} t_{\rm stop}$,  and $t_{\rm stop}$ is the stopping time that measures how fast the particle adjusts its momentum to the surrounding gas. The headwind prefactor that measures the disk gas pressure gradient  is given by    
\begin{equation}
  \eta = - \frac{1}{2}  \frac{\partial \ln P }{\partial \ln r} \left(\frac{H_{\rm g}}{ r} \right)^2 = C_{\eta} h_{\rm g}^2,
  \label{eq:eta}
\end{equation}
where  $h_{\rm g} {=} H_{\rm g}/r$ is the disk aspect ratio, $H_{\rm g}$ is the gas scale height,  $C_{\eta}$ is an order unit factor related to the disk structure. Combing \eq{vgas2} and \eq{vpeb},  the outward movement of dust particles requires  
\begin{equation} 
  \frac{ \alpha}{\taus} > \frac{4C_{\eta}}{3} \frac{r_{\rm exp}}{r - r_{\rm exp}}.
  \label{eq:criterion}
\end{equation}
   
The final size that particles can reach depends on the growth pattern.  We consider two circumstances: particles are either in fragmentation-limited or bouncing-limited regimes\cite{Guttler2010}.  In the fragmentation-limited growth regime the highest Stokes number of particles is given by $\tausF {\equiv} v_{\rm F}^2/ \alpha c_{\rm s}^2$,  where $c_{\rm s}{=}H_{\rm g}\Omegak$ is the gas sound speed and $v_{\rm F}{=}1 \ \rm m/s $ is the adopted fragmentation threshold speed, related to the composition and size of the monomer grains\cite{Blum2008,Gundlach2015}. The turbulent speed of the dust particles is given by $ \Delta v_{\rm t}{=}\sqrt{\alpha \taus} c_{\rm s}$. Inserting it into \eq{criterion}, we obtain the criterion of particles' outward drift in the fragmentation regime:
\begin{equation}  
  \alpha > \frac{v_{\rm F}}{c_{\rm s}} \sqrt{\frac{4C_{\eta}}{3} \frac{r_{\rm exp}}{r - r_{\rm exp}}}.
  \label{eq:criterion_F}
\end{equation}
As can be seen from the above equation, a high $\alpha$ and hot disk promotes the outward movement of dust particles.
   
On the other hand, some laboratory experiments find that the growth of the particles is stalled at around $0.1{-}1$mm  when the  bouncing effect becomes dominant\cite{Guttler2010,Zsom2010}. In this work we assume that  particles  in the bouncing-limited growth regime reach a radius of $ 0.3 \rm \  mm$.  The particles' Stokes number  reads  
\begin{equation}  
  \tausB  =\sqrt{2 \pi} \frac{\rho_{\bullet} R_0}{  \Sigma_{\rm g}},
  \label{eq:sizeStokes}
\end{equation}   
where $\rho_{\bullet}$ is the internal density of the dust particles,  and $R_0$ is the radius of the bouncing particles. Inserting \eq{sizeStokes} into \eq{criterion}, we obtain the criterion of particles' outward drift in the bouncing regime:
\begin{equation} 
  \alpha > \frac{4 \sqrt{2 \pi}C_{\eta}}{3} \frac{\rho_{\bullet}R_0}{\Sigma_{\rm g}}  \frac{r_{\rm exp}}{r - r_{\rm exp}}.
  \label{eq:criterion_B}
\end{equation}

We assume that dust particles have already reached the maximum sizes at the beginning of numerical integration for the above two cases.  This is valid as long as the growth timescale is shorter than the viscous spreading timescale. The growth timescale from micron-sized grains to mm-sized particles that initiate substantial radial drift is given by $t_{\rm grow}{\sim}\xi (Z \Omega_{\rm K})^{-1} $, where $\xi{\sim}10$ and $Z{=}10^{-2}$ is the canonical dust-to-gas ratio, while the viscous timescale is $t_{\nu }{\sim} (\alpha h_{\rm g}^2 \Omega_{\rm K})^{-1} $. It is easy to calculate that $t_{\rm grow}{\ll} t_{\nu}$ for standard disk parameters. 
     
The turbulent velocity is isotropic when the MRI is the predominant disk turbulence source\cite{Johansen2005}. The MRI is not able to operate when the ionization level is low Hence, a layered accretion model has also been proposed, where the turbulence in the midplane is strongly damped and accretion is mainly through an ionized, active surface layer\cite{Gammie1996}. We account for both these two models: one assumes a uniform disk turbulence (constant $\alpha$ everywhere), and the other incorporates the effect of the dead zone.  In the latter model, $\alpha$ -- which sets the global viscous evolution of the disk -- is much higher than the turbulent level $\alphat$ in the quiescent midplane layer. The ionization relates to the column density of the disk gas, which increases with orbital distance. We adopt a simple prescription for $\alphat$ in the dead zone model,
\begin{equation}  
  \alphat = \alpha_{\rm D} + (\alpha_{\rm A} -  \alpha_{\rm D}) \times \frac{r-r_{\rm in}}{r_{\rm out} - r_{\rm in}},
\end{equation}
where $\alpha_{\rm D}$ and $\alpha_{A}{=}\alpha$ are the low and high boundary of turbulent viscosity parameter and $r_{\rm in}$ is the inner disk radius where the disk temperature is $1000$ K.  
    
\noindent {\bf Parameter study}

We investigate how the radial drift of dust particles influenced by different parameters in both the uniform turbulence and dead zone models.  The particles are either in the fragmentation-limited or in the bouncing-limited growth regimes.  The parameter setups are provided in Tables $1$, while the simulation results are depicted in \Fgs{migrationF}{migrationB}.  For the size of the particles limited by the fragmentation threshold,  we find in \Fg{migrationF} that particles maintain their outward movement for a less time when the disk is less turbulent, the temperature is lower, the initial disk size is larger,  and the fragmentation threshold velocity is higher.  Meanwhile,  the outward-then-inward drifting particles that eventually enter the asteroid belt have a larger size when the disk is less turbulent, the temperature is lower, the initial disk size is larger, the disk is more massive, the fragmentation threshold velocity is higher and/or the particles are more porous with a lower internal density.  

We note that the fragmentation velocity is a crucial factor in determining the duration of the NC-CC separation,  since the Stokes number of the particles directly correlates to $v_{\rm F}^2$. We find that very large pebbles form when the fragmentation threshold is high, (\eg , $v_{\rm F}{=}3 \rm  \ m/s$ in \Fg{migrationF}h). Such particles quickly decouple from the diffusing gas and drift into the inner disk within less than $1$ Myr. It is therefore difficult to generate the NC-CC separation if the fragmentation velocity is very high.

For the size of the particles limited by the bouncing barrier, we find in \Fg{migrationB} that particles beyond certain disk radii (${\gtrsim} 20$ AU) drift in a convergent manner.  Further out particles have high Stoker numbers that already decouple from the disk gas to move inward,  while particles with slightly inner orbits still couple with the outwardly spreading gas. The bouncing-limited size is also varied at $0.1$, $0.3$, and $1$ mm.  When the dust particles reach a smaller size,  they can couple to the expanding gas for a longer time, resulting in a longer NC-CC separation. Similarly, porous dust particles with a lower internal density have a lower Stokes number compared to those with a compact structure. Therefore,  the porous dust particles  can maintain the outward expansion for a longer time.  To summarize, in order to separate the two disk reservoirs for more than $2{-3}$ Myr,  a lower bouncing-limited size or particles with a lower internal density are preferred.

In order to analyze the results statistically,  we also perform a high number of Monte Carlo sampling simulations by varying two disk parameters at each time. The CC material injection radius $r_{\rm CC}$ is uniformly chosen from $5$ to $35$ AU, the initial disk size $r_{\rm d}$ is uniformly chosen from $10$ to $40$ AU, the disk temperature at $1$ AU ranges from $120$  to $400\rm \ K$,  and $\alpha$ is log-uniformly selected from $10^{-3}$ to $10^{-2}$ in the uniform turbulent model or $\alpha_{\rm D}$  from  $10^{-4}$ to $10^{-3}$ in the dead zone model. The particles in the bouncing-limited regime are assumed to be $0.6$ mm in diameter. The results are given in \Fg{parameterFB}.

The duration of the NC-CC separation in the expanding disk model is strongly influenced by the properties of the protoplanetary disk. \Fg{parameterFB} shows parameter dependencies of the arrival time of CC particles in the inner disk. For the explored range of parameters, the time for CC particles reaching the inner $1$ AU disk region spans from $10^{5}$ to $10^{7}$ yr, with a later entering time for a larger $r_{\rm CC}$. Overall, the inner boundary of CC material larger than $20$ AU is needed for the corresponding drift time of ${>}$ a few Myr. The drift timescale of particles beyond $r_{\rm CC}$ sensitively depends on the initial disk size, since it directly correlates to places where disk gas can diffuse inward/outward ($r_{\rm exp}$).

\noindent {\bf Early disk evolution}

Our study separates the disk-forming stage from the disk-evolution stage.  We define the former phase as the period in which gas from the parent molecular cloud accretes onto the central protostellar core and the surrounding circumstellar disk, while in the latter phase this process has come to an end. The disk is then approximated as a closed and isolated system where gas accretion onto the host star is paired with an expanding outer disk, in line with conventional disk theory and observations\cite{Hartmann1998}. In the main paper, we show that this disk evolution phase is sufficient to explain the observed isotopic dichotomy. There is, however, a significant degree of uncertainty on the earlier disk building phase 
 \cite{Nordlund2014,Kuffmeier2017,Lebreuilly_2021}.

Here we show that our assumptions on the disk formation stage are in line with theoretical and observational constraints.
Firstly, disks  likely form rapidly: a conventional Bonnor-Ebert sphere collapses within a free-fall timescale of $T_{\rm ff} {= }\sqrt{r_{\rm BE}^3/GM_{\star}} {\sim}0.4$ Myr, where we take the core mass to be a solar mass and a critical radial extend of $r_{\rm BE}$ to be  ${\approx}0.1$ pc.  Although some early condensates like CAIs may form within this early period,  most solids form in the disk evolution phase (\Fg{timetable}). 
This estimated free-fall timescale is compatible with the range of disk formation timescales measured in the state-of-the-art magneto-hydrodynamical simulations\cite{Nordlund2014, Kuffmeier2017,Lebreuilly_2021}.

It is less clear how exactly mass is deposited onto the disk during this disk-forming phase. In the classical picture, not considering magnetic fields,
material will collapse inside-out when the infalling material conserves its angular momentum.  The cloud-collapsing front expands outward,  with high angular momentum material deposited onto the disk at a later time, e.g. \cite{Appelgren2020}.

Current magneto-hydrodynamic (MHD) simulations similarly argue for inside-out formation without a strong imprint of magnetic braking: ideal MHD simulations typically show a strong and changing misalignment between the total angular momentum and magnetic field vectors of young disks, when taking into account the turbulent nature of the giant molecular cloud \cite{Kuffmeier2017}. Therefore, disks undergo relatively isotropic accretion -- with strong local and temporal variation -- of gas at $\sim$\,$50$\, AU that results in a radial mass flow in the disk midplane\cite{Nordlund2014, Kuffmeier2017}.
We therefore assume that, within the infall phase, the gas only accretes inward. After $t_{\rm infall}$, the disk evolution switches to the nominal situation described in Section \textbf{Spreading disk model}.  Recently, it has been argued that magnetic braking and vertical accretion play a larger role \cite{Hennebelle_2020, Lebreuilly_2021}. However, a statistical analysis of these results, starting from the molecular cloud level, shows that disk properties for ideal and non-ideal MHD (including ambipolar diffusion) simulations appear to be largely similar, but in the latter case disks tend to be less massive ($<0.1 \ $ M$_{\star}$) \cite{Lebreuilly_2021}. Regardless, the current mass and radial extent of the Solar System points towards it not forming from such a potential population of small, low-mass disks, but instead from an intermediate-sized disk of size $>50$\, AU. Moreover, outer Solar System solids with a low degree of thermal processing, like comets and CI chondrites, were more likely delivered radially from the outer regions of the collapsing cloud. This is in line with the model presented here with an outer -- more pristine -- CC component. Delivering the bulk material closer to the host star would give a potentially excessive degree of thermal processing.

\noindent {\bf Jupiter's growth and migration}

We adopt the state-of-the-art pebble-driven planet formation model\cite{Liu2019}.  The key equations  for Jupiter's growth and migration are recapitulated as follows. 

The  pebble accretion efficiency in the $2$D and $3$D limits read
 \begin{equation}
\begin{split}
\varepsilon_\mathrm{PA,2D} & 
 = \frac{0.64 (1 + \taus^2)}{2\taus \eta + (1 + \taus^2) u_{\rm g}/v_{\rm K}} \sqrt{ \frac{M_{\rm p} }{M_\star}   \frac{\Delta v}{ v_{\rm K} }  \taus  } 
\label{eq:eps-2D}
\end{split}
\end{equation}
and 
\begin{equation}
\label{eq:eps-3D}
\begin{split}
\varepsilon_\mathrm{PA, 3D} & =  \frac{0.78 }{\left[2 \eta + (1 + \taus^2) (u_{\rm g}/v_{\rm K})/\taus \right]h_\mathrm{peb}} \frac{M_{\rm p}}{ M_\star}.
 \end{split}
\end{equation}
We take into account of the gas flow velocity in the above equations,  $M_{\rm p}$ and $M_{\star}$ are the masses of the planet and the star,  and $ \Delta v$ is the relative velocity between the planet and the pebble. 

The pebble mass flux ($\dot M_{\rm peb}$) is assumed to be attached to the gas flow with a constant mass flux ratio such that $\xi {\equiv}\dot M_{\rm peb}/\dot M_{\rm g}$.  Particles of very low Stokes number are well-coupled to the disk gas. The gas and dust particles drift together with the same velocities, and therefore $\xi$ remains equal to the initial disk metallicity.  When the Stokes number is high, pebbles drift faster than the disk gas.  In this case,  in order to maintain a constant flux ratio, $\Sigma_{\rm peb}/\Sigma_{\rm g}$ becomes lower than the nominal disk metallicity and the pebble surface density gets reduced. 
The mass accretion onto the planetary core can be written as  
\begin{equation}
\frac{d M_{\rm p, PA}}{dt} =  \varepsilon_\mathrm{PA}  \dot M_{\rm peb} =  \varepsilon_\mathrm{PA} \xi \dot M_{\rm g}.
\end{equation}

The core mass growth is terminated when the planet reaches the pebble isolation mass,
 \begin{equation}
   \begin{split}
 M_{\rm iso} = & 25 \left( \frac{h_{\rm g}}{0.05} \right)^3 \left( \frac{M_{\star}}{\Ms} \right) \left[ 0.34 \left(  \frac{ -3}{ {\rm log} \alphat}  \right)^4 + 0.66 \right] \\
  & \left[ 1-   \frac{ \partial  {\rm ln } P /  \partial {\rm ln} r +2.5} {6}        \right] \Me.
 \label{eq:m_iso}
 \end{split}
 \end{equation}

After reaching the pebble isolation mass,  the planet starts rapid gas accretion 
  \begin{equation}
 \dot M_{\rm p, g} = \min \left[ \left(\frac{d M_{\rm p, g}}{dt}\right)_{\rm KH} ,\left(\frac{d M_{\rm p, g}}{dt}\right)_{\rm Hill}, \dot M_{\rm g}    \right],
 \label{eq:gas_acc}
 \end{equation}  
The first term  in \Eq{gas_acc} corresponds to the Kelvin-Helmholtz contraction 
  \begin{equation}
  \begin{split}
 \left(\frac{d M_{\rm p, g}}{dt}\right)_{\rm KH} =10^{-5} \left( \frac{\Mp }{10 \Me} \right)^4 \left( \frac{\kappa_{\rm env}}{1 \rm \ cm^2g^{-1}}  \right)^{-1} \rm \  M_{\oplus}\,yr^{-1},
 \label{eq:gas_KH}
 \end{split}
 \end{equation}
 where  $\kappa_{\rm env}$ is the opacity in of the planetary envelope. 
 
The second term in \Eq{gas_acc} indicates how much gas the in the planet Hill is able to get accreted by the planet, 
  \begin{equation}
\left(\frac{d M_{\rm p,  g}}{dt}\right)_{\rm Hill} = \frac{f_{\rm acc}}{3 \pi} \left( \frac{R_{\rm H}}{H_{\rm g}} \right)^2 \frac{\dot M_{\rm g}}{\alpha} \frac{\Sigma_{\rm gap}}{\Sigma_{\rm g}},
\label{eq:gas_hill} 
\end{equation}
where $R_{\rm H} {=} (M_{\rm p}/3M_{\star})^{1/3} r$ is the Hill radius of the planet, and $\Sigma_{\rm g}$ and $\Sigma_{\rm gap}$ are the unperturbed gas surface density and the surface density at the bottom of the gap. The relative depth of the gap can be given by 
\begin{equation}
 \frac{\Sigma_{\rm g}}{\Sigma_{\rm gap}} =1 + 0.04 \left( \frac{M_{\rm p}}{ M_{\star}}\right)^{2} \left(\frac{1}{h_{\rm g}} \right)^{5}  \left(\frac{ 1}{\alpha} \right).
\label{eq:sig_gap}
\end{equation}
We also define the gap opening mass $M_{\rm gap}$ when $\Sigma_{\rm gap}$ is reduced by $50\%$ compared to the unperturbed value.

Briefly,  the Kelvin-Helmholtz contraction is the dominant gas accretion channel when the planet mass is low.  When  $(d M_{\rm p g}/ dt)_{\rm KH}{>}(d M_{\rm p g}/dt)_{\rm Hill}$, the gas accretion is limited by the amount of gas within the planet's Hill sphere.  Once the planet becomes sufficiently massive,  the available gas is eventually determined by the global disk accretion flow.   In this work,  $\kappa_{\rm env}$ and  $f_{\rm acc}$ are the two variables that determine the timescale of gas accretion.

The planet interacts with sounding disk gas,  inducing orbital migration. 
Such a planetary migration rate is given by 
\begin{equation}
\dot r =  \dot r_{\rm I}/ \left[ 1 + \left(\frac{M_{\rm p}}{M_{\rm gap}}\right)^2 \right].
\label{sec:migration rate}
\end{equation}
This formula summarizes for both the low-mass planet in the type I and  massive planet in the type II torque regimes.  
The type I migration rate reads 
  \begin{equation}
\dot r_{\rm I} = -f_{\rm I} \left( \frac{\Mp}{M_{\star}} \right)  \left( \frac{\Sigma_{\rm g} r^2}{ M_{\star}}  \right) h_{\rm g}^{-2} v_{\rm K},
\label{eq:type_I}
\end{equation} 
 where $f_{\rm I}{=}2$ is adopted from Eq.(8) of Ref.\cite{Kley2012} approximately for the migration prefactor in a local isothermal disk. 

 We start the growth track of a protoplanet with $1\Me$ and  $t_0$,$r_0$, $\xi$, $f_{\rm acc}$ and $\kappa_{\rm env}$ are the parameters that can be varied to generate the current Jupiter's mass and location. One example is illustrated in \Fg{dichotomy}c,  where $t_0 {=}10^{5}$ yr, $r_0{=}45 $ AU, $\xi {= }0.04$, $\kappa_{\rm env}{=}0.05 \rm \ g/cm^{2}$ and $f_{\rm acc}{=}0.9$.  The other parameters are: $v_{\rm F}{=}1.5 \rm \ m/s$,  $r_{\rm d}{=}8$ AU, $C_{\rm d}{=}0.14 \ M_{\odot}$,   $T_{\rm irr0}{=} 130$ K, $\alpha_{\rm A}{=}8 \times10^{-3}$ and $\alpha_{\rm D}{=} 10^{-3}$ (dead zone model).

\noindent {\bf  Pebble accretion for asteroids on eccentric and inclined orbits }

We explain how the inclination and eccentricity affect an asteroid's growth by pebble accretion.  The orbits of the planetesimals are excited through their mutual gravitational interactions. The random velocities of massive embryos remain low, whereas the eccentricities/inclinations of small asteroids become high. There are two consequences. First, the orbits of asteroids are lifted away from the dust midplane layer due to increased inclinations, resulting in fewer pebbles able to be accreted. Second, the pebble-asteroid encounter time becomes short owing to their high relative velocity. When the encounter time is shorter than the stopping time of pebbles, the pebble accretion efficiency drops greatly. Both effects are more profound for lower-mass planetesimals.  

\Fg{PA} shows the mass growth of asteroids varying with their eccentricities and inclinations. The parent-bodies of ordinary and enstatite chondrites form in the NC reservoir at $t{=}2$ Myr.  However, even though $100 \ \Me$ CC dust particles drift across their orbits over the next $1$ Myr,  these asteroids accrete an insignificant amount to reverse their isotopic signatures as long as their eccentricities/inclinations are beyond a few times  $10^{-3}$.

\clearpage
\noindent {\bf \large Supplementary Figures and Tables} \\

 \begin{table}[!h]
\begin{threeparttable}
    \centering
    \caption{ Model setups for the parameter study.  The radius of particles is either chosen to be $0.1$ (fiduical), $0.3$ or $1$ mm in the bouncing-limited growth regime or calculated from their Stokes number of $\tau_{\rm s, F}{=}v_{\rm F}^2/(\alpha_{\rm t} c_{\rm s}^2) $ in the fragmentation-limited growth regime where $v_{\rm F}{=} 1$ (fiducial), $0.3$ or $3 \rm  \ m/s$. }
    \begin{tabular}{c|c|c|c|c|c|c|c|c|c|c|c}
        \hline
        \hline
  $\rm  Run$  & \multicolumn{2}{c}{$\rm dead \ zone$} & $\rm uniform $    &   $T_{\rm irr0} $  &  $ C_{\rm d}$  & $v_{\rm F}$ &  $R_{\rm peb}$  &  $ r_{\rm d}$ &  $\rho_{\bullet}$\\
     & $\alpha({=}\alpha_{\rm A})$  & $\alpha_{\rm D}$ & $\rm  turbulence \ \alpha$ &    $ [\rm K]$  &  $   [\Ms]$  & $\rm [m/s]$  & $\rm [mm]$  &  $ \rm [AU]$ &  $ \rm [gcm^{-3}]$\\
       \hline
    $\rm fiducial$ & $ 10^{-2}$ & $5 \times 10^{-4}$ & $5 \times 10^{-3}$ &  $130$ &  $0.14$  & $1$ & $0.3$ & $ 10$ & $1.5$\\
  $\rm  low \ \alpha$  & $ 10^{-3}$ & $5 \times 10^{-4}$ & $ 10^{-3}$ &  $130$ &  $0.14$  &  $1$ & $0.3$   & $ 10$ & $1.5$\\
  $\rm  hot \ disk$  & $10^{-2}$ & $5 \times 10^{-4}$ & $5 \times 10^{-3}$ &  $280$ &  $0.14$  &  $1$ & $0.3$   & $ 10$ & $1.5$\\
    $\rm low \ mass \ disk$  & $ 10^{-2}$ & $5 \times 10^{-4}$ & $5 \times 10^{-3}$ &  $130$ &  $0.014$  &  $1$ & $0.3$   & $ 10$ & $1.5$\\
    $\rm large \ size \ disk$   & $10^{-2}$ & $5 \times 10^{-4}$ & $5 \times 10^{-3}$ &  $130$ &  $0.14$  &  $1$ & $0.3$   & $ 20$ & $1.5$\\
     $\rm porous \ particle$ & $ 10^{-2}$ & $5 \times 10^{-4}$ & $5 \times 10^{-3}$ &  $130$ &  $0.14$  &  $1$ & $0.3$  & $ 10$ & $0.15$ \\
     $\rm high \  v_{\rm F}$  & $10^{-2}$ & $5 \times 10^{-4}$  & $5 \times 10^{-3}$ &   $130$ &  $0.14$  & $3$ & $\setminus$ & $ 10$ & $1.5$\\
      $\rm low \  v_{\rm F}$  & $10^{-2}$ & $5 \times 10^{-4}$  & $5 \times 10^{-3}$ &   $130$ &  $0.14$  & $0.3$ & $\setminus$ & $ 10$ & $1.5$\\
      $\rm large \ particle $  & $10^{-2}$ & $5 \times 10^{-4}$  & $5 \times 10^{-3}$ &   $130$ &  $0.14$ & $\setminus$  & $1$  & $ 10$ & $1.5$\\
      $\rm small \ particle $  & $10^{-2}$ & $5 \times 10^{-4}$  & $5 \times 10^{-3}$ &   $130$ &  $0.14$ & $\setminus$ & $0.1$  & $ 10$ & $1.5$\\
   \hline
   \hline
     \end{tabular}
       \end{threeparttable}
    \label{tab:model}
\end{table} 
\newpage

 \begin{figure}
 \includegraphics[width=12cm]{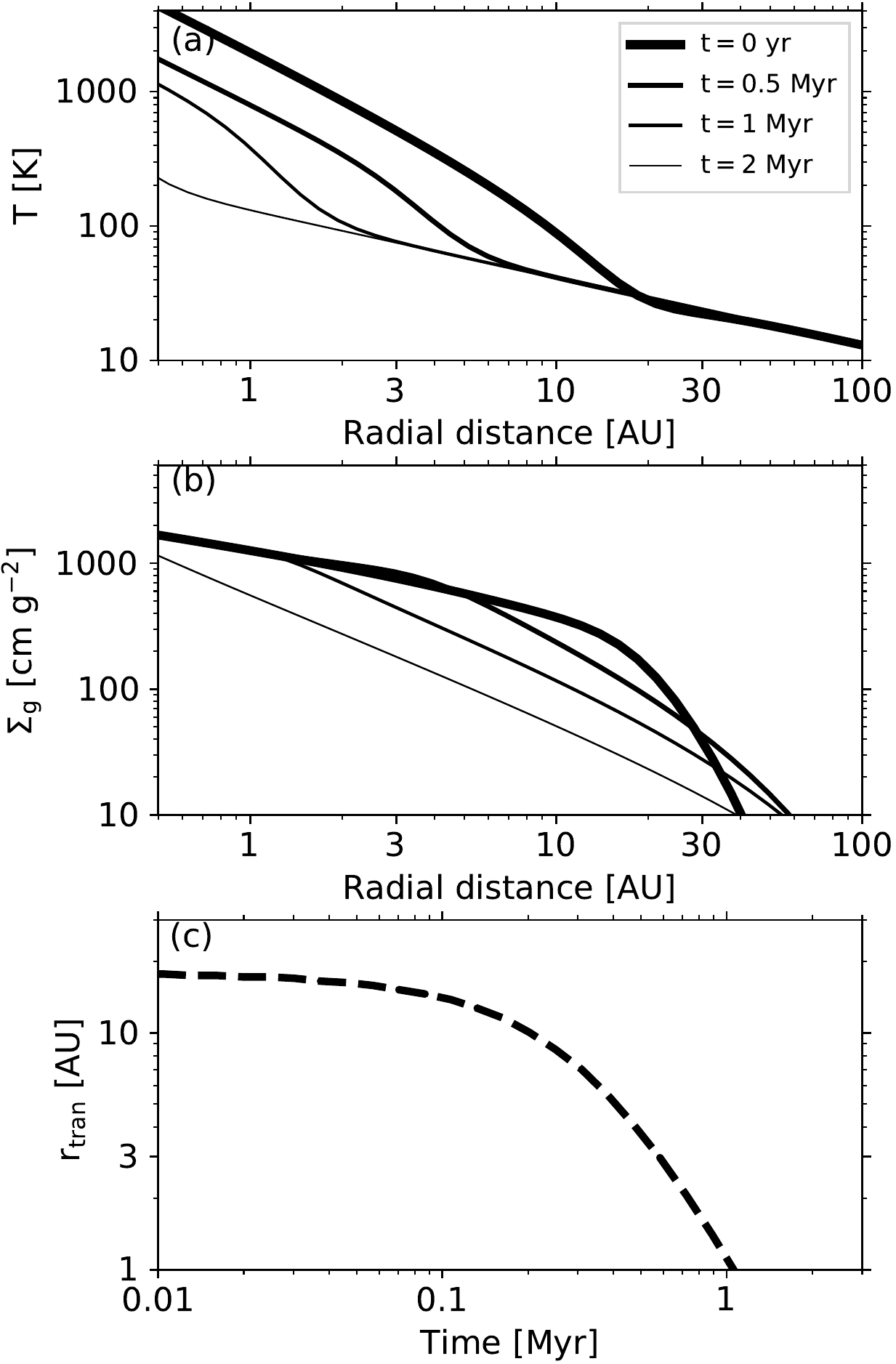}
\caption{\textbf{Two-component structure for a spreading disk.} Panel (a): disk temperature as a function of radial distance.   Panel (b): gas surface density as a function of radial distance.  The thickness of line represents  different time of $t{=}0$ yr,  $0.5$ Myr,  $1$ Myr and $2$ Myr, respectively.  Panel (c): time evolution of the transition radius.  The parameters are adopted from the fiducial run with a uniform turbulence model in Table $1$. 
} 
\label{fig:diskmodel}
\end{figure}
\newpage

\begin{figure}
  \includegraphics[width=1\linewidth]{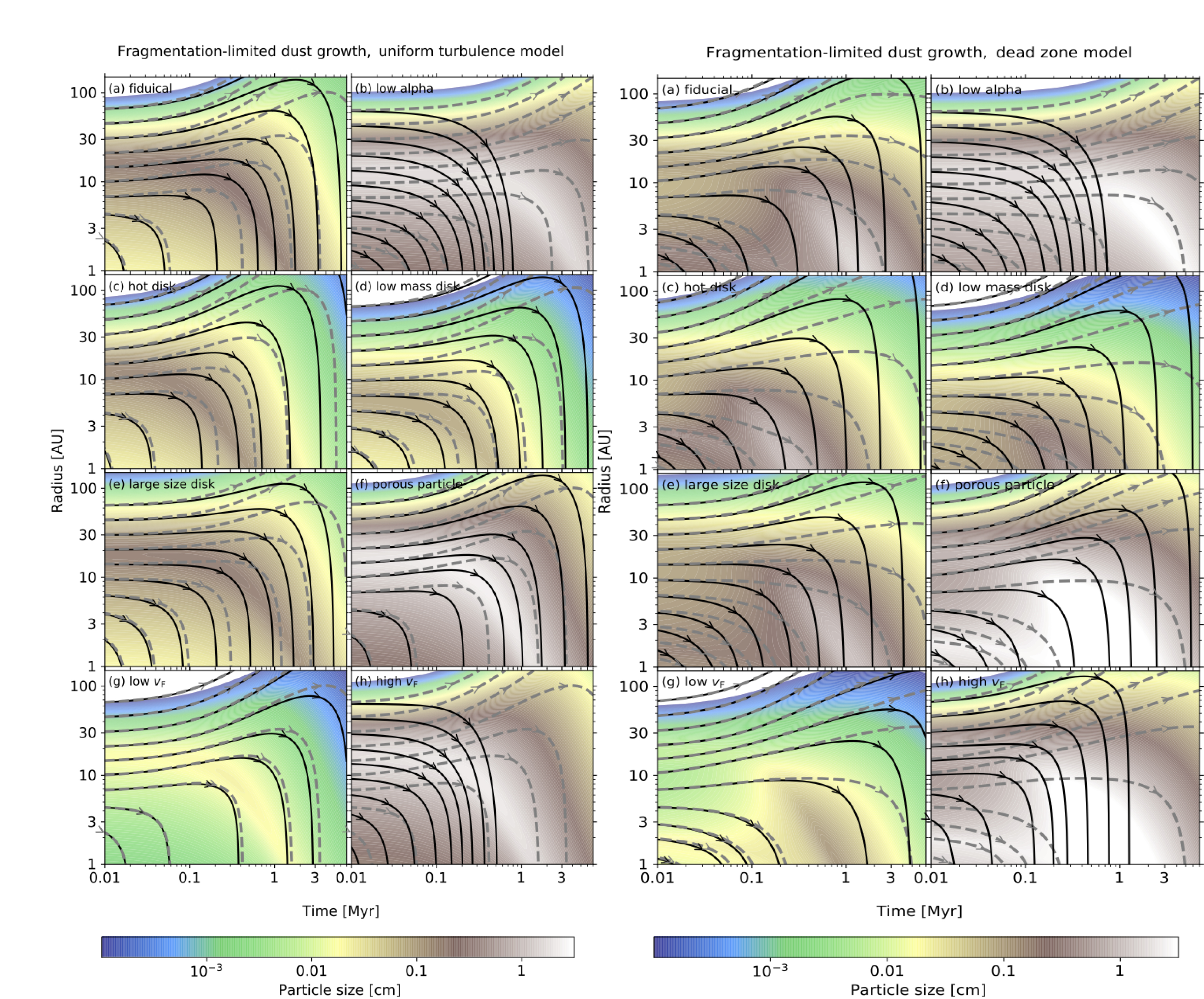}
\caption{\textbf{ Parameter study on the radial drift of dust particles with fragmentation-limited sizes in the uniform turbulence (left)  and dead zone (right) disk model.}  The black curves trace the radial drift of dust particles, while the grey dashed lines indicate the gas spreading at the same initial disk locations.  The color refers to the size of the particle at the corresponding time and disk location.  The model setups are presented in Table $1$.
} 
\label{fig:migrationF}
\end{figure}
\newpage

 \begin{figure}
   \includegraphics[width=1\linewidth]{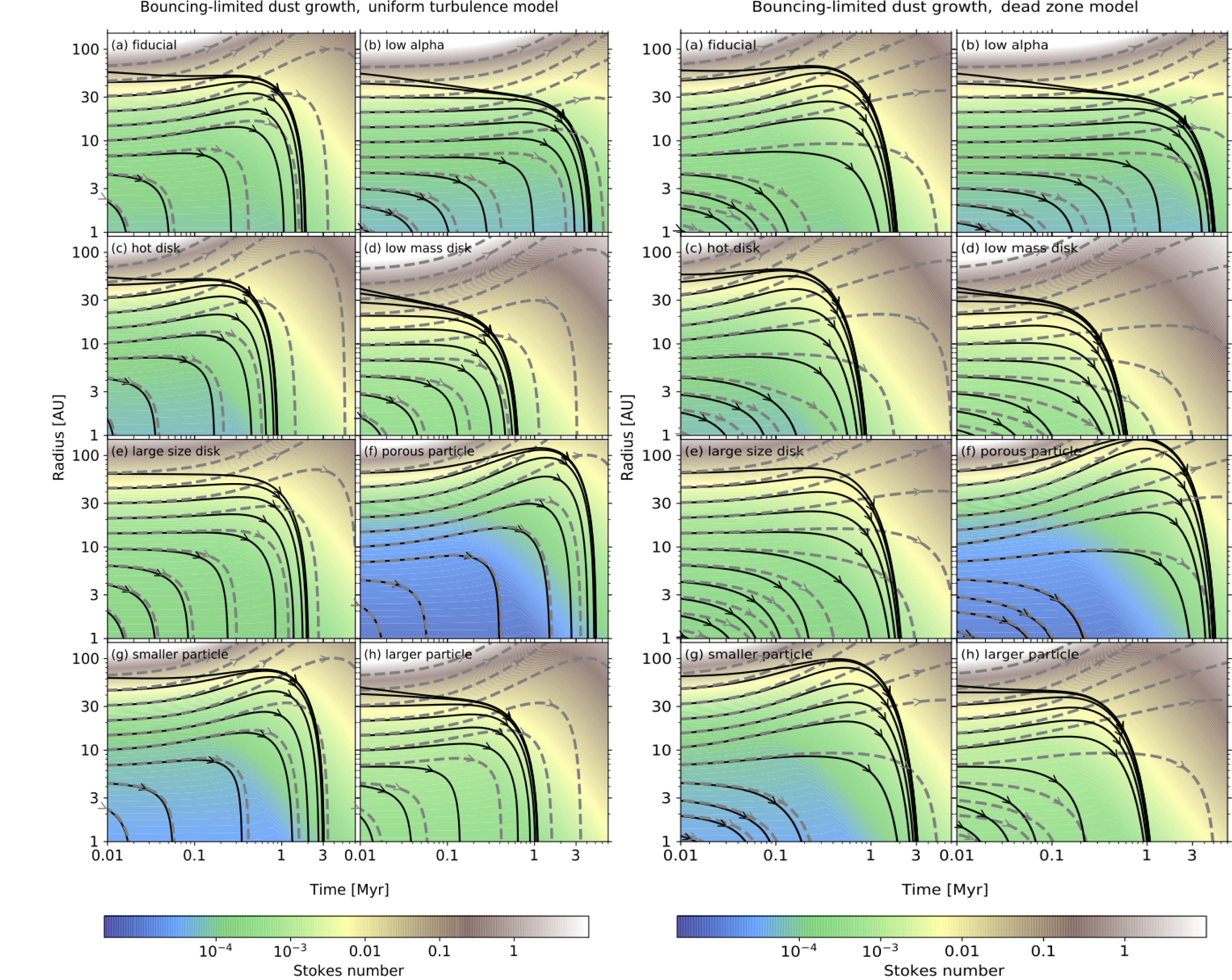}
\caption{\textbf{ Parameter study on the radial drift of dust particles with bouncing-limited sizes in the uniform turbulence (left) and dead zone (right) disk model.}  The black curves trace the radial drift of dust particles, while the grey dashed lines indicate the gas spreading at the same initial disk locations.  The color refers to the Stokes number of the particle at the corresponding time and disk location.  The model setups are presented in Table $1$.
} 
\label{fig:migrationB}
\end{figure}

\newpage

\begin{figure}
\centering
\includegraphics[width=14cm]{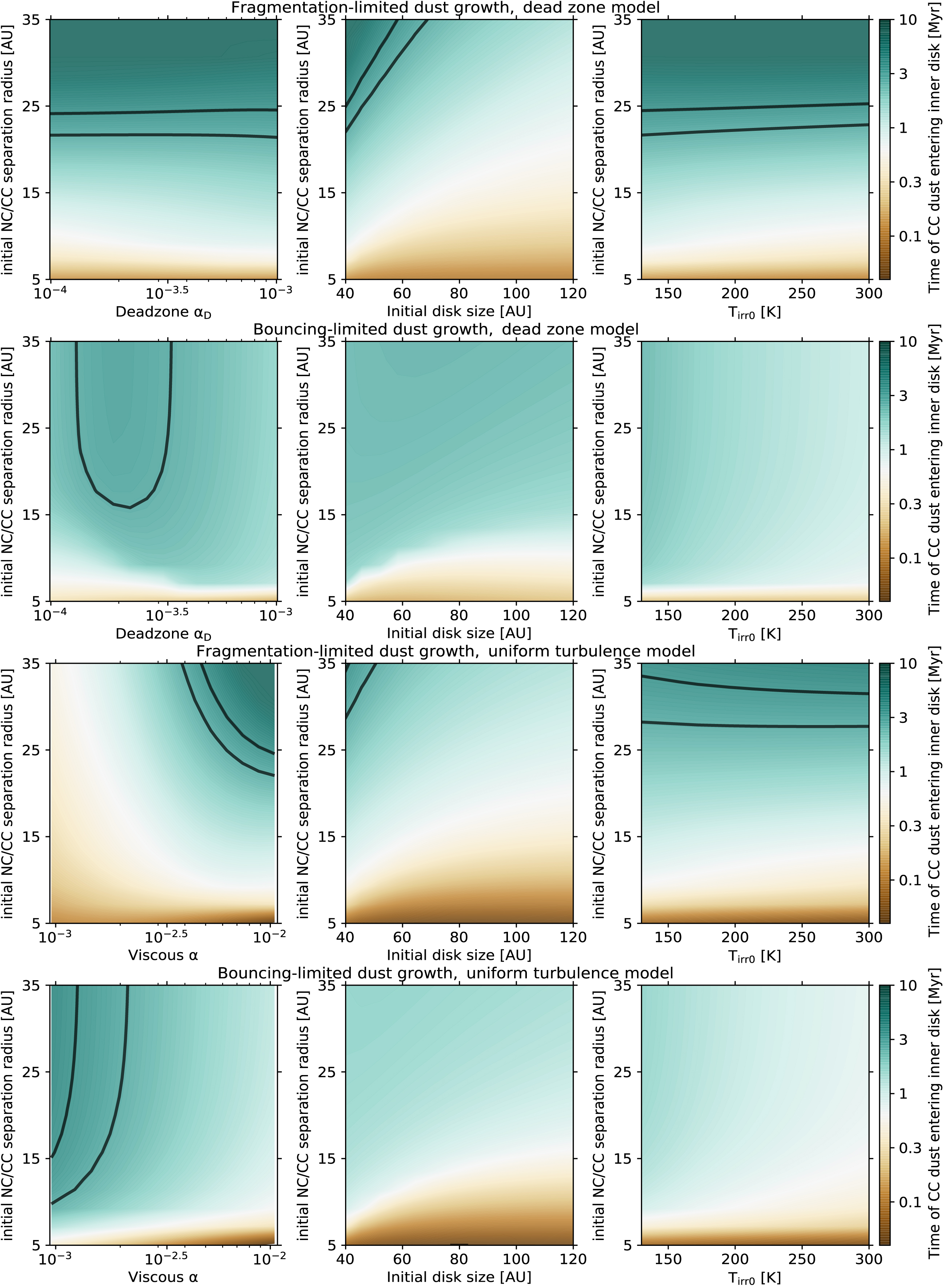}
\caption{\textbf{Monte Carlo simulation outcomes as a function of disk parameters.} The dead zone and uniform turbulence models with particles in the fragmentation and  bouncing-limited regimes are illustrated in from top to bottom. The color corresponds to the time of CC dust particles entering the inner disk region of $r{<}1$ AU. The circles with black edgecolor refer to the systems that CC particles entering the terrestrial planet formation region during $3{-}4$ Myr after the CAI formation.  We fix the other model parameters and vary the initial CC material injection radius vs midplane $\alpha$ (left),  initial disk size (middle) and $T_{\rm irr0}$ (right), respectively.  The default parameters are $\alpha_{\rm D}{ =} 5\times 10^{-4}$, $\alpha_{A}{=}10^{-2}$ (for the dead zone model) or $\alpha{=} 5\times 10^{-3}$ (for the uniform turbulence model), $r_{\rm d} {=}10$
AU,  $v_{\rm F}{=}1 \rm \ m/s$,  $T_{\rm irr0}{=}130$ K and  $C_{\rm d}{=}0.14 \ M_{\odot}$, respectively.}
\label{fig:parameterFB}
\end{figure}
\newpage

\begin{figure}[h!]
\includegraphics[width=16cm]{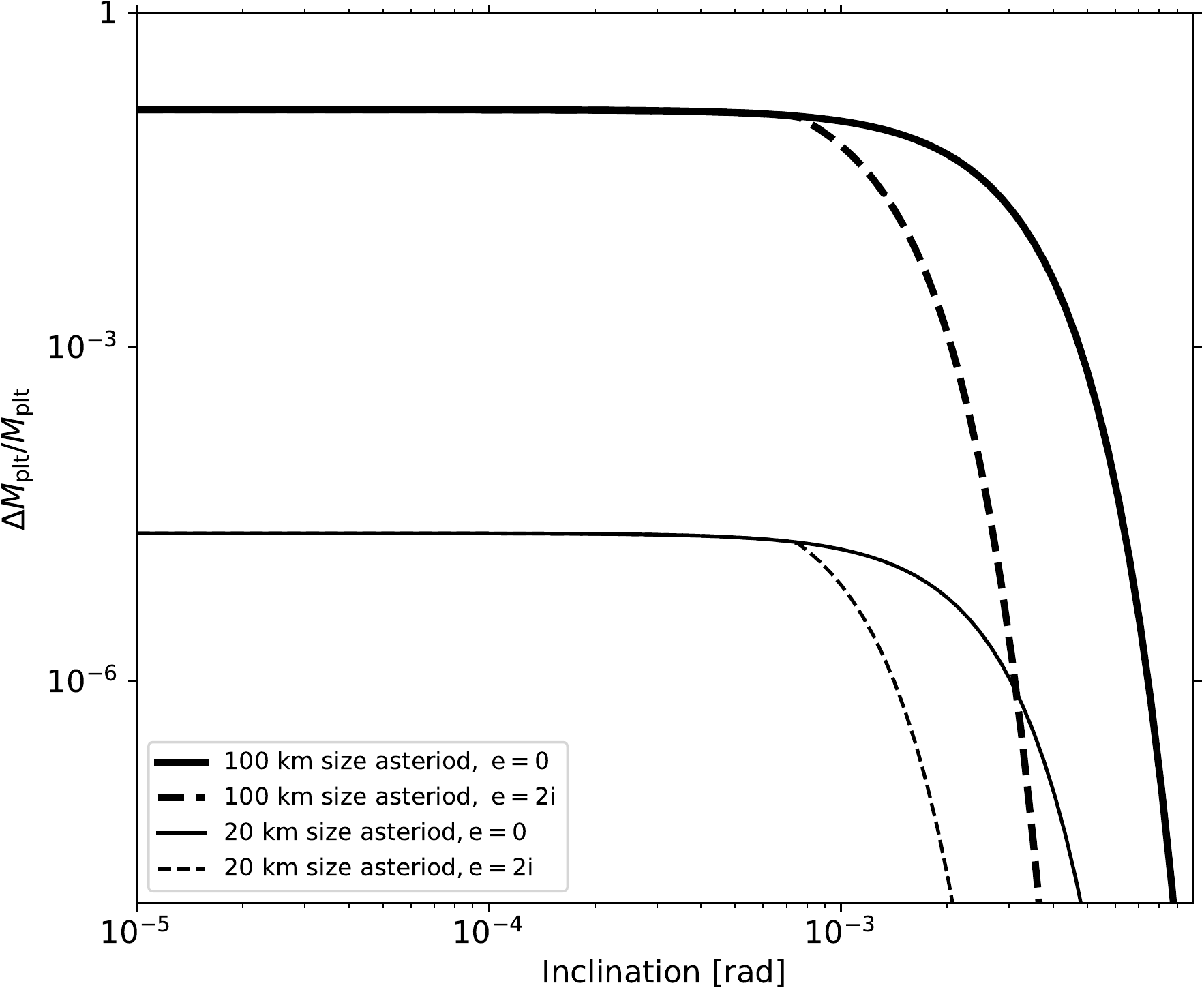}
\caption{\textbf{Planetesimal mass growth as a function of inclination.} The initial sizes of planetesimals are $100$ km (thick) and $20$ km (thin), whereas their orbits are either circular (solid) or with eccentricities two times of inclinations (dashed). The disk is adopted to be the fiducial setup of the uniform turbulence model and the size of particles is bouncing-limited to be one millimeter.  The planetesimals are assumed to form at $t{=}2$ Myr and the following mass growth by pebble accretion is accounted for when $100 \ \Me$ pebbles drift bypass their orbits. The pebble accretion prescription is adopted from Refs\cite{Liu2018,Ormel2018}.     
}
\label{fig:PA}
\end{figure}

\clearpage

\bibliography{reference}
\bibliographystyle{ScienceAdvances.bst}

\clearpage
\noindent{\bf \large Acknowledgments}

\noindent{\bf Funding} \\
B.~L. is supported by the start-up grant of the Bairen program from Zhejiang University, National Natural Science Foundation of China (No.12173035 and 12111530175), and the Swedish Walter Gyllenberg Foundation.  A.~J. acknowledges funding from the European Research Foundation (ERC Consolidator Grant 724687-PLANETESYS), the Knut and Alice Wallenberg Foundation (Wallenberg Scholar Grant 2019.0442), the Swedish Research Council (Project Grant 2018-04867), the Danish National Research Foundation (DNRF Chair Grant DNRF159) and the G\"oran Gustafsson Foundation. M.~B. acknowledges funding from the Carlsberg Foundation (CF18 1105) and the European Research Council (ERC Advanced Grant 833275-DEEPTIME).  T.~H. acknowledges funding from the Independent Research Fund Denmark (DFF8021-00350B).

\noindent{\bf Author contributions} \\
A.~J., B.~L. and M.~L. contribute equally to initiating the idea and drafting the manuscript.  B.~L. performed numerical simulations. All authors contributed to analyzing and discussing the results, their implications and in finalizing the manuscript.

\noindent{\bf Competing interests} \\
The authors declare that they have no competing interests.

\noindent{\bf Data and materials availability} \\
All data needed to evaluate the conclusions in the paper are present in the paper and the Supplementary Materials.

\end{document}